\documentclass[aps,prl,numerical,twocolumn]{revtex4-1}
\usepackage{graphicx}
\usepackage{bm}
\usepackage{amssymb}
\usepackage{multirow}
\usepackage[colorlinks=true,linkcolor=blue,citecolor=blue,urlcolor=blue]{hyperref}

\begin{document}

\title{The exponential-nonexponential transition in quantum decay as a phenomenon of interference in time domain of the decaying particle with itself}
\author{Gast\'on Garc\'{\i}a-Calder\'on}
\email[email:]{gaston@fisica.unam.mx}
\author{Roberto Romo}
\altaffiliation[Permanent address:]{ Facultad de Ciencias, Universidad Aut\'onoma de Baja California, 22800 Ensenada, Baja California, Mexico}
\affiliation{Instituto de F\'{\i}sica, Universidad Nacional Aut\'onoma de M\'exico, 04510 Ciudad de M\'exico, Mexico}

\date{\today}
\begin{abstract}
By using an exact analytical non-Hermitian approach in terms of resonance (quasinormal) states we express the decaying wave function as the sum of exponential and nonexponential decaying solutions to the time-dependent Schr\"odinger equation. We show that the exponential-nonexponential transition of decay at long times represents physically a process of interference in time domain of the decaying particle with itself.
\end{abstract}

\maketitle

\textit{Introduction.} In 1928, in the early days of quantum mechanics, Gamow imposed on physical grounds outgoing (radiative) boundary conditions to the solutions of the Schr\"odinger equation to describe $\alpha$-decay in radioactive nuclei \cite{gamow28a,gamow28b,gamow49}. This led to complex energy eigenvalues and to the derivation of the exponential decay law $\exp(-\Gamma t/\hbar)$ for the evolving probability density, where the decay rate $\Gamma$ corresponds to the imaginary part of the complex energy eigenvalue. Around the end of the fifties of last century Khalfin \cite{khalfin58} pointed out that in decaying  systems where the energy spectra is bounded by below, i.e., $E \in (0, \infty)$, which includes most physical systems of interest, it follows due to a theorem by Paley and Wiener \cite{wiener34} that the exponential decay law cannot be valid at all times. Khalfin considered in his analysis the survival probability, which yields the probability that at time $t$ the decaying particle remains in its initial state. He was able to show that this quantity exhibits, in addition to a purely decaying exponential behavior that follows by assuming a complex pole located on the energy plane, an integral contribution that  behaves at long times as an inverse power of time. Most subsequent work on this subject \cite{levy59,newton61,winter61,goldberger64b,terentev72,chiu77,peres80,berger82}, has been strongly influenced by the work by Khalfin.

One should notice, however, that the result obtained by Khalfin is based on a mathematical argument and  hence it does not provide a physical mechanism to understand  the exponential-nonexponential transition. An approach to deal with this question was considered by  Fonda and Ghirardi \cite{fonda72}  who  following the work by Ersak \cite{ersak69} argued that the physical mechanism for the deviation from exponential decay law at long times is a partial regeneration process of the initial state caused by rescattering of the decayed states \cite{fonda72}. However, as discussed below, exact model calculations show that the proposed mechanism yields a negligible contribution to the exact nonexponential behavior at long times, and more importantly, it does not provide a description of the exponential-nonexponential transition.

The present work rests on an exact analytical non-Hermitian formulation of quantum tunneling decay \cite{gc10} which involves the complex poles of the propagator and the resonance (quasinormal) states to the problem to address the issue of the physical mechanism of the exponential-nonexponential transition at long times. We demonstrate that the decaying wave function may be written as the sum of exponential and nonexponential decaying wave functions. The latter involves the propagation of almost vanishing values of the wave number and as time evolves eventually interferes with the exponential decaying terms which refer to wave components close to resonance energy. This interference yields the exponential-nonexponential transition. We show that physically it corresponds to a phenomenon of interference in time domain of the decaying particle with itself.

It is worth mentioning that the failure to find deviations of the exponential decay law  at long times in radioactive nuclei \cite{norman88,son98}, contributed to the widespread view that nonexponential decay contributions were beyond experimental reach or even to the alternative explanation that the interaction of the decaying system with the environment would enforce exponential decay at all times \cite{ghirardi78,parrot02}. However, the experimental verification in recent times of short-time deviations from exponential decay \cite{raizen97} and the quantum Zeno effect \cite{itano90,raizen01} together with the measurement of the deviations from exponential decay law at long times in  organic molecules in solution, that exhibited distinct  inverse power behaviors in time \cite{monk06}, have demonstrated that nonexponential decay is an observable quantum effect.

The formulation considered here refers to the full Hamiltonian to the problem and hence it differs from approaches where the Hamiltonian is separated into a part corresponding to a closed system and a part responsible for the decay which is usually treated to some degree of perturbation theory, as in the work by Weisskopf and Wigner on the  exponential decay of an excited atom interacting with a quantized radiation field \cite{wigner30a} or in studies concerning the deviation of exponential decay in these systems \cite{knight77}.

\textit{Formalism.} The eigenfunctions associated with complex energy eigenvalues, the resonance states, increase beyond the interaction region exponentially with distance implying that the usual rules concerning normalization and completeness do not apply. For these reasons the approach by Gamow has been considered a phenomenological no fundamental approximation for the description of decaying systems. However,  modern developments of the formalism of resonance states have solved in a consistent fashion the above issues \cite{gc10}. It has been shown that this non-Hermitian approach  yields exactly the same results for the time evolution of decay that a Hermitian approach based on continuum wave solutions for generic exactly solvable models \cite{gcmv12,gcmv07,gcmv13}.

Here, we briefly recall  the relevant aspects of the derivation of the decaying wave solution for a single particle confined initially within the internal region of a spherically symmetric real potential with the condition, imposed on physical grounds, that it vanishes beyond a distance, i.e., $V(r)=0$ for $r>a$. We choose natural units $\hbar=2m=1$ and for simplicity of the discussion and without loss of generality we refer to $s$ waves. The solution to the time-dependent Schr\"odinger equation may be written in terms of the retarded Green's function $g(r,r';t)$ of the problem as \cite{gc10},
\begin{equation}
\Psi(r,t)=\int_0^a {\! g(r,r^\prime;t)\Psi(r',0)\,\mathrm{d}r^\prime},
\label{1s}
\end{equation}
where $\Psi(r,0)$ stands for an arbitrary initial state which is confined within the internal interaction region.
The retarded time-dependent Green's function $g(r,r';t)$ is the relevant quantity to study the time evolution of the initial state. It
may be evaluated  by a Laplace transformation into the complex wave number plane $k$ aimed to exploit the analytical properties of the outgoing Green's function to the problem $G^+(r,r';k)$ \cite{gcp76,gc10},
\begin{equation}
g(r,r';t)={1 \over 2 \pi i} \int_{c_\circ} G^+(r,r\,';k) {\rm e}^{-i k^2t} \,2kdk,
\label{74}
\end{equation}
where $c_\circ$ refers to the Bromwich contour which corresponds to an hyperbolic contour along the first quadrant of the $k$ plane.
A consequence of the condition that the potential vanishes after a distance is that $G^+(r,r\,';k)$ may be extended analytically to whole complex $k$ plane where it has an infinite number of complex poles distributed in a well known manner \cite{newtonchap12}.
Resonance states and complex energy poles are intimately related. Resonance states are solutions to the radial Schr\"odinger equation
$[E_n-H]u_n(r)=0$ obeying outgoing (radiative) boundary conditions $[du_n(r)/dr]_{r=a}=i\kappa_n(a)$, where $E_n=\kappa_n^2=\mathcal{E}_n-i \Gamma_n/2$. Here, $\mathcal{E}_n$ stands for the resonance energy of the decaying particle and $\Gamma_n$ for the corresponding resonance width. As is well known, the longest lifetime sets up the time scale of the decay process.
Resonance states may be also obtained from the residues at the complex poles $\{\kappa_n\}$ of the outgoing Green's function which also provides its  normalization condition \cite{gcp76,gc10}, namely, $\int_0^a u_n^2(r) dr + i{u_n^2(a)}/{2\kappa_n}=1$. It is worth mentioning that resonance states satisfy flux conservation \cite{gc10}.
The above considerations allow for the rigorous derivation of the resonance expansion of the outgoing Green's function \cite{gc10},
\begin{equation}
G^+(r,r\,';k) = \sum_{n=-\infty}^{\infty} \frac {u_n(r)u_n(r\,')}{2\kappa_n(k-\kappa_n)}, \quad  (r,r')^{\dagger} \leq a
\label{9x}
\end{equation}
where the above sum includes the resonance states $u_{-n}(r)$ and poles $\kappa_{-n}$ located on the third quadrant of the $k$ plane which are related to those located on the fourth quadrant by symmetry relations that follow from time reversal invariance: $\kappa_{-n}= -\kappa_n^*$ and $u_{-n}(r) = u^{*}_n(r)$ \cite{rosenfeld61,gc10}; the notation $(r,r')^{\dagger} \leq a$ means that the point $r = r' = a$ is excluded in the above expansion, since otherwise it diverges.

The representation of $G^+(r,r\,';k)$ given by (\ref{9x}) satisfies the closure relation \cite{gcmv12,gc10},
\begin{equation}
{\rm Re} \left\{\sum_{n=1}^{\infty} u_n(r)u_n(r\,')\right\}=\delta(r-r\,'),\quad  (r,r')^{\dagger} \leq a,
\label{9y}
\end{equation}
and the sum rules \cite{gc10}, with $(r,r')^{\dagger} \leq a$,
\begin{equation}
\sum_{n=-\infty}^{\infty} \frac{u_n(r)u_n(r\,')}{\kappa_n}=0, \quad \sum_{n=-\infty}^{\infty} u_n(r)u_n(r\,')\kappa_n=0.
\label{9yy}
\end{equation}
One may also write the resonance expansion of the Green's function given by (\ref{9x}), using the identity
$1/[2\kappa_n(k-\kappa_n)] \equiv  1/2k[1/(k-\kappa_n) +1/\kappa_n]$ and (\ref{9yy}) as,
\begin{equation}
G^+(r,r\,';k) = \frac{1}{2k}\sum_{n=-\infty}^{\infty} \frac {u_n(r)u_n(r\,')}{k-\kappa_n}. \quad  (r,r')^{\dagger} \leq a
\label{9xa}
\end{equation}

The evaluation of $g(r,r';t)$ as a resonance-state expansion involving the poles of $G^+(r,r';k)$ may be obtained by distinct deformations of the contour $c_\circ$. One of them leads to an integral extending along the full real $k$ axis \cite{gc10}. Then
substitution of (\ref{9xa}) into (\ref{1s}) allow us to write the time-dependent decaying wave function as \cite{gcmv12,gc10},
\begin{equation}
\Psi(r,t)=\sum_{n=-\infty}^{\infty}
\left \{ \begin{array}{cc}
C_nu_n(r)M(y^\circ_n), & \quad  r \leq a \\[.4cm]
C_nu_n(a)M(y_n), & \quad r \geq a,
\end{array}
\label{b6}
\right.
\end{equation}
where the sums run over the full set of poles, the coefficients $C_n$ are defined by,
\begin{equation}
C_n=\int_0^a \Psi(r,0) u_n(r) dr,
 \label{3c}
\end{equation}
and the functions $M(y_n)$, the so called Moshinsky functions, are defined as \cite{gc10},
\begin{equation}
M(y_n)=\frac{i}{2\pi}\int_{-\infty}^{\infty}\frac{{\rm e}^{ik(r-a)}{\rm e}^{-ik^2t}}{k-\kappa_n}dk=
\frac{1}{2}{\rm e}^{(ir^2/4 t)} w(iy_n),
\label{16c}
\end{equation}
with $y_n={\rm e}^{-i\pi /4}(1/4t)^{1/2}[(r-a)-2 \kappa_nt]$,
and the function $w(z)=\exp(-z^2)\rm{erfc(-iz)}$ in (\ref{16c}) stands for the Faddeyeva-Terent'ev or complex error function \cite{abramowitzchap7} for which there exist efficient computational tools to calculate it \cite{poppe90}. The argument $y_n^{\circ}$ of the functions $M(y_n^0)$ in (\ref{b6}) is that of $y_n$ with $r=a$, namely,
\begin{equation}
y^\circ_n=-{\rm e}^{-i\pi /4}\kappa_nt^{1/2}.
\label{16ci}
\end{equation}
Assuming that the initial state $\Psi(r,0)$ is normalized to unity, it follows from the closure relation (\ref{9y}) that,
\begin{equation}
{\rm Re}\sum_{n=1}^\infty \left\{ C_n \bar{C}_n\right\}= 1,
\label{9z}
\end{equation}
where ${\bar C}_n$ follows by taking the conjugate of $\Psi(r,0)$ in (\ref{3c}).
Equation (\ref{9z}) indicates that ${\rm Re}\,\{C_n{\bar C}_n\}$ cannot be interpreted as a probability, since in general it is not a positive definite quantity. Nevertheless, one may see that it represents the `strength'  or `weight' of the initial state in the corresponding resonant state. One may see the coefficients  ${\rm Re}\,\{C_n{\bar C}_n\}$ as some sort of quasi-probabilities.

The solution $\Psi(r,t)$ for $r \leq a$,  given by the first equation in (\ref{b6}), is the relevant ingredient to calculate the survival probability, as discussed in \cite{gcmv12,gcmv07,gc10}. For $r \geq a$, the solution $\Psi(r,t)$, given by the second equation in (\ref{b6}),  describes the propagation of a single decaying particle along the external region. This has been discussed in Refs. \cite{gcmv12,gcmv13,gc10}.

The exponential and nonexponential explicit behavior of $\Psi(r,t)$ for $r \leq a$ may be achieved
by using the symmetry relations  mentioned above among the poles located on the third and fourth quadrants on the $k$ plane, $\kappa_{-n}=-\kappa_n^*$ and correspondingly for the resonance states, $u_{-n}(r)=u_n^*(r)$. As a result one may write $\Psi(r,t)$ for $r \leq a$ as,
\begin{equation}
\Psi(r,t)= \sum_{n=1}^\infty [C_nu_n(r)M(y_n^0)+{\bar C}_n^* u_n^*(r)M(y_{-n}^0)].
\label{i2}
\end{equation}
One then may utilize a property of the functions $M(y_n^0)$ that establishes that  $M(y_n^0) = \exp (-i\kappa_n^2t) -M(-y_n^0)$, provided that $\pi/2 < \arg \,(y_n^0) < 3\pi/2$ \cite{moshinsky52,gc10}. This is in fact the case for resonance poles with $\alpha_n > \beta_n$, the so called \textit{proper} resonance poles. In such a case, the arguments of both $M(-y_n^0)$ and $M(y_{-n}^0)$, satisfy $-\pi/2 < \arg \,(y_n^0) < \pi/2$, and hence do not exhibit an exponential behavior. As a result, one may write (\ref{i2}) as,
\begin{equation}
\Psi(r,t)= \Psi_{e}(r,t) + \Psi_{ne}(r,t),\qquad r \leq a
\label{i2a}
\end{equation}
where $\Psi_{e}(r,t)$ corresponds to the sum of exponential decaying terms
\begin{equation}
\Psi_{e}(r,t)= \sum_{n=1}^\infty C_nu_n(r)e^{-i\mathcal{E}_n t}e^{-\Gamma_n t/2},
\label{i3}
\end{equation}
and $\Psi_{ne}(r,t)$ stands for the nonexponential contribution,
\begin{equation}
\Psi_{ne}(r,t)= -\sum_{n=1}^\infty [ C_nu_n(r) M(-y_n^0)-{\bar C}_n^*u_n^*(r)M(y_{-n}^0)].
\label{i4}
\end{equation}
The expressions for $\Psi_{e}(r,t)$ and  $\Psi_{ne}(r,t)$, given by (\ref{i3})  and  (\ref{i4}), satisfy the time-dependent Schr\"odinger equation.

Here we shall analyze the exponential-nonexponential transition, without loss of generality, for the time evolution of the survival probability $S(t)=|A(t)|^2$, where the survival amplitude $A(t)$ is defined as,
\begin{equation}
A(t)=\int_0^a \Psi^*(r,0) \Psi(r,t)\,dr.
\label{i1}
\end{equation}
Hence, using (\ref{i3}) and (\ref{i4}) one may write (\ref{i1}) as
\begin{equation}
A(t)=A_{e}(t) + A_{ne}(t)
\label{i5}
\end{equation}
where
\begin{equation}
A_{e}(t)=\sum_{n=1}^\infty C_n{\bar C}_ne^{-i\mathcal{E}_n t}e^{-\Gamma_n t/2},
\label{i6}
\end{equation}
and,
\begin{equation}
A_{ne}(t)= -\sum_{n=1}^\infty [ C_n{\bar C}_n M(-y_n^0)-(C_n{\bar C}_n)^*M(y_{-n}^0)].
\label{i7}
\end{equation}
One sees immediately by inspection of (\ref{i6}) that the exponential decaying behavior of the survival amplitude  corresponds to the sum of distinct resonance energies weighted by expansion coefficients that fulfill (\ref{9z}).
It turns out that for large values of the argument, the $M$ functions in (\ref{i7}) exhibit an asymptotic expansion that goes as $M(z_q) \sim 1/z_q -1/z_q^3 +...$, with $z_q=-y_n^0$ or $y_{-n}^0$ \cite{gc10, abramowitzchap7}. The leading term in these expansions, in view of the first expression in (\ref{9yy}) and (\ref{16ci}), vanishes exactly and hence,
\begin{equation}
A_{ne}(t) \approx  -i \eta\,  {\rm Im} \left [\sum_{n=1}^{\infty}\frac{C_n{\bar
C}_n}{\kappa_n^3} \right ] \frac{1}{t^{\,3/2}},
\label{i8}
\end{equation}
with $\eta=1/(4\pi i)^{1/2}$.

In order to establish which energies (wave numbers) contribute to the nonexponential expression given by (\ref{i8}), it is convenient to evaluate $g(r,r';t)$  by closing the contour $c_0$ in (\ref{74}) in a different fashion, namely by considering a line $45^\circ$ off the real axis that may be evaluated by deforming  $c_0$  and using the Theorem of residues to get explicitly the exponential decaying contributions plus an integral term along that line \cite{gcmv07,gc10}. The integral term may be evaluated by the steepest descents method. The essential point is that the saddle point is at $k=0$ and  making a Taylor's expansion of $G^+(r,r';k)$ around that point one may write \cite{gcmv07,gc10},
\begin{eqnarray}
&&g(r,r\,';t) \approx \sum_{n=1}^{\infty} u_n(r)u_n(r')e^{-i \mathcal{E}_nt} e^{-\Gamma_n t/2} \nonumber \\ [.3cm]
&& -i\eta \left \{ \frac{\partial}{\partial k} G^+(r,r';k)\right\}_{k=0}
\frac{1}{t^{3/2}}. \quad  (r,r')^{\dagger} \leq a
\label{74a}
\end{eqnarray}
Equation (\ref{74a}) shows beyond any doubt that the nonexponential contribution arises from almost vanishing values of the wave number $k$ and hence of the energy. This behavior is known \cite{newtonchap19,peres80}, but as far as we know, has never been used to investigate the physical mechanism for the exponential-noexponential transition.  One may use (\ref{9x}) to expand the second term in (\ref{74a}) in terms of resonance states, to obtain,
\begin{equation}
-i \eta\, {\rm Im}\left\{\sum_{n=1}^{\infty}\frac{u_n(r)u_n(r\,')}{\kappa_n^3} \right\} \frac{1}{t^{\,3/2}}, \quad
(r,r')^{\dagger} \leq a
\label{74b}
\end{equation}
It follows then using  (\ref{1s}) and (\ref{i5}), that (\ref{74b}) leads exactly to the expression of the nonexponential survival amplitude at long times given by  (\ref{i8}). In an analogous way, the first term in (\ref{74a}) corresponds exactly to (\ref{i6}).

From (\ref{i5}) the survival probability near the  exponential-nonexponential transition reads,
\begin{equation}
S(t) \approx |A_{e}(t)|^2 + |A_{ne}(t)|^2 + 2 {\rm Re} [A^*_{e}(t)A_{ne}(t)],
\label{i5a}
\end{equation}
where $A_{e}(t)$ and $A_{ne}(t)$ are given by (\ref{i6}) and (\ref{i8}). Depending on the parameters of the potential, there is a time $t_0$ where the first and second terms in (\ref{i5a}) are necessarily of the same order of magnitude. Around  $t_0$,  the  last term in (\ref{i5a}) reads,
\begin{eqnarray}
&&2 {\rm Re} [A^*_{e}(t)A_{ne}(t)] \approx  - 2\, {\rm Re} \times \nonumber \\ [.4cm]
&& \left \{ \left [\sum_{n=1}^\infty C_n{\bar C}_ne^{-i\mathcal{E}_n t}e^{-\Gamma_n t/2} \right ]^*
\left [i \eta {\rm Im} \sum_{n=1}^{\infty}\frac{C_n{\bar C}_n}{\kappa_n^3\, t^{\,3/2}} \right ] \right \}.
\nonumber \\ [.4cm]
\label{i8a}
\end{eqnarray}
The above expression exhibits analytically the interference in time domain of the exponential and the nonexponential contributions to the decay process in the transition region, which represents the main result of this work.

Since the coefficients $C_n{\bar C}_n$ satisfy the closure relation given by (\ref{9z}), in practice, depending on the initial state a finite number of terms are needed to evaluate (\ref{i8a}). It is straightforward to see that the above considerations hold also for the time evolution of the probability density $|\Psi(r,t)|^2$.

Let us now, refer to the physical mechanism proposed by Fonda and Ghirardi to  produce the deviations from the exponential decay law. The expression obtained by Ersak \cite{ersak69} may be written as,
\begin{equation}
I(t,T)= A(t+T)-A(t)A(T),
\label{i8b}
\end{equation}
where $I(t+T)$ represents, according to Fonda and Ghirardi \cite{fonda72}, the physical mechanism for producing deviations from the exponential decay law that follows as a consequence that the decayed states  reconstruct partially the initial state through a process of rescattering. We show below by evaluating exactly the right-hand side of (\ref{i8b}) that contrary to the claim by Fonda and Ghirardi, $|I(t,T)|^2$,  yields a negligible contribution to nonexponential decay.

\begin{figure}
\begin{center}
\includegraphics[width=0.95\linewidth]{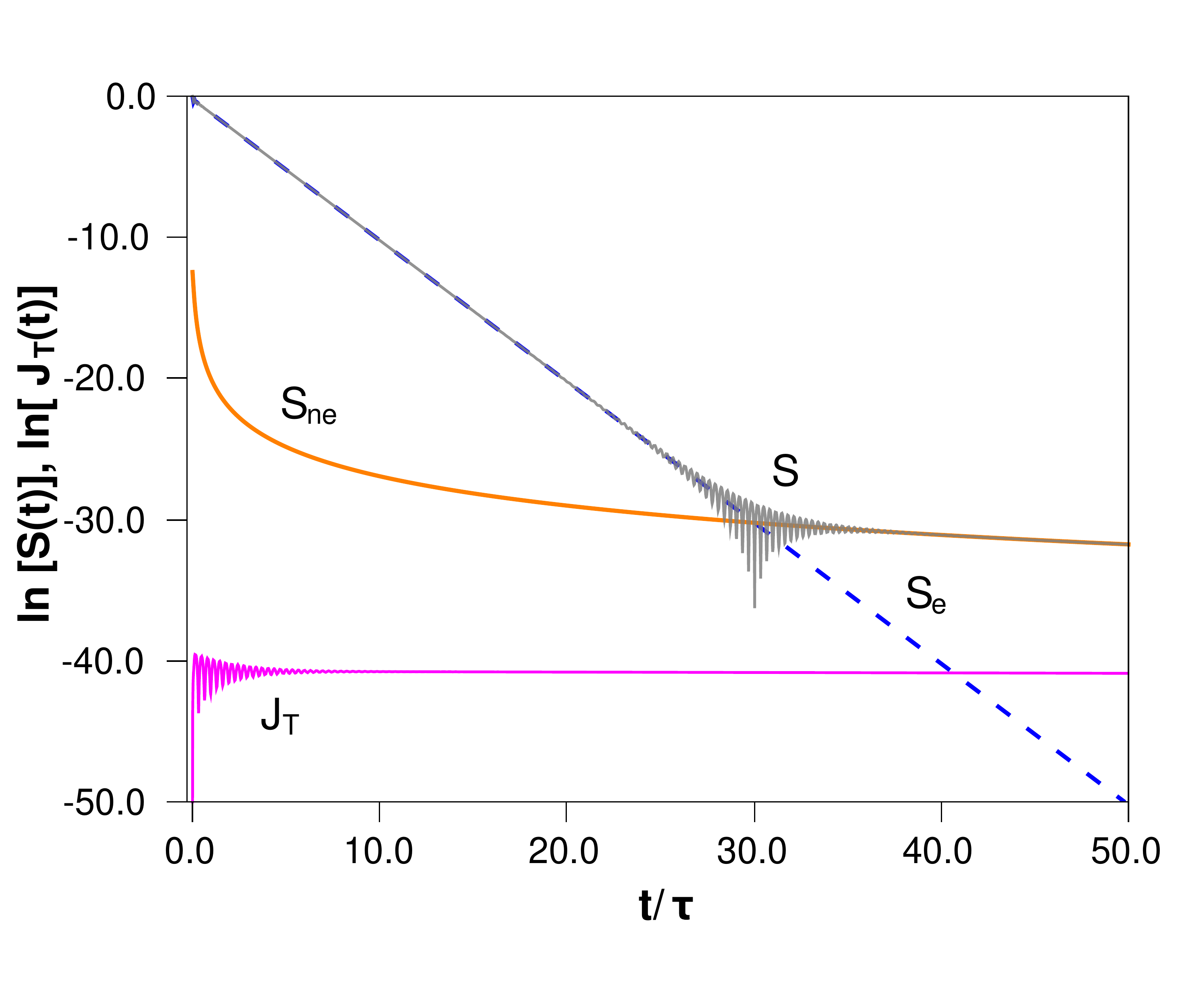}
\caption{Natural logarithm of the survival probability $S(t)$ (solid grey line) as a function of the dimensionless time $t/\tau$. Also shown are the exponential, $S_e(t)=|A_{e}(t)|^2$, and nonexponential, $S_{ne}(t)=|A_{ne}(t)|^2$ contributions, and the Ersak term $J_T(t)=|I(t,T|^2$, with $T=1000$, a value within the nonexponential regime. Notice that $J_T(t)$ is much smaller than the nonexponential contribution of $S(t)$. See text.}
\label{figure1}
\end{center}
\end{figure}
\textit{Model.} As an example, let us consider the \textit{barrier shell potential}, which consists of a well of width $w$, followed by a rectangular potential barrier of width $b$, so $w+b=a$. The system parameters are: barrier height, $V=30.0$; well width, $w=1.0$; and barrier width, $b=0.3$.
The resonance state solutions to the Schr\"odinger equation obeying outgoing boundary conditions
with the usual continuity conditions at the distinct interfaces of the potential lead to the equation whose solution yields, following known procedures \cite{nussenzveig59,cgc10a,vertse14}, the $\kappa_n$'s to the problem.

The resonance parameters for the first resonance are: the resonance energy ${\mathcal E}_1=6.84$ and the resonance width $\Gamma_1=0.352$. As initial state we use a quantum-box state $\Psi(r,0)=\sqrt{2/w}\sin{(\pi r/w)}$ placed at the center of the quantum well.

Figure~\ref{figure1} shows a typical survival probability graph along the exponential and long-time regimes (solid grey line), which exhibits in particular the exponential-nonexponential transition which is described analytically by (\ref{i8a}). In the present example, the initial state overlaps strongly with the first resonance state of the problem $u_1(r)$, namely, ${\rm Re}\, C_1^2 =0.899$. This implies, in view of (\ref{9z}), that using the first resonance state and the corresponding pole, is a good approximation to describe the time evolution of decay of the problem. This allows us to make use of a formula for the exponential-nonexponential transition time derived for a single resonance level \cite{gcrr01}, which in lifetime units reads $\tau_0=5.41 \ln (R)+12.25$, with $R=\mathcal{E}_1/\Gamma_1=19.4$ and hence $\tau_0 = 28.29$, which is in very good agreement with the calculation.
Figure \ref{figure1} also displays the exponential, $\ln S_e(t)$, and nonexponential, $\ln S_{ne}(t)$, contributions given by (\ref{i6}) and (\ref{i8}).
Also shown in Fig. \ref{figure1} is an exact calculation of the Ersak term $J_T(t)=|(I(t,T)|^2$ with $T=1000$, a value within the nonexponential regime of the survival probability. One sees that $\ln J_{1000}(t)$ yields a negligible contribution to the exact nonexponential contribution $\ln S_{ne}(t)$ and does not describe the exponential-nonexponential transition.

\textit{Concluding remarks.} The main contribution of this work is to provide, using an exact analytical non-Hermitian description, the physical mechanism for the exponential-nonexponential transition in the time evolution of quantum decay as arising from the interference in time domain of the exponential and nonexponential contributions of the decaying wave function. In our view this explanation exhibits a hitherto unknown feature of an exact nonstationary solution to the Schr\"odinger equation and therefore invites  to explore both possible physical consequences of this mechanism and its implication on foundational issues of quantum mechanics. Our approach follows a line of inquiry that explores the possibility to extend the standard formalism of quantum mechanics to incorporate in a fundamental fashion the present non-Hermitian treatment of the Hamiltonian to the system \cite{gcmv12,gcv19}.

\textit{Acknowledgments.}  G.G-C. acknowledges financial support of DGAPA-UNAM-PAPIIT grant IN105216, Mexico and  R.R.  acknowledges financial support from PRODEP under the program \textit{Apoyo para Estancias Cortas de Investigaci\'on}, Convocatoria 2018. R.R. also thanks Instituto de F\'{\i}sica of UNAM for its hospitality.
\end{document}